\documentclass[12pt]{article}
\usepackage{a4}
\usepackage{epsfig,url}
\usepackage{caption}
\usepackage{lineno}

\begin{document}
\title{German Beams\\ The Story of Particle Accelerators in Germany}
\author{Volker Ziemann, Uppsala University, Uppsala, Sweden}
\date{\today}
\maketitle
\begin{abstract}\noindent
  Even though many of the experiments leading to the standard model of particle physics
  were done at large accelerator laboratories in the US and at CERN, many exciting developments
  happened in smaller national facilities all over the world. In this report we highlight
  the history of accelerator facilities in Germany.
\end{abstract}
In {\em Beams}\footnote{V. Ziemann, {\em Beams---the Story of Particle Accelerators
    and the Science they Discover,} Copernicus Books, Springer 2024.} I describe the
co-evolution of particle accelerators and the standard model of elementary particles.
The majority of these experiments were done at the large laboratories in the US and
at CERN. There are, however, so many other accelerators doing important experiments
that it would be a shame not to briefly talk about them as well. In the coming
sections I will tell the story of accelerators in Germany.
\subsection*{Prehistory}
%
%
The first particle accelerator in Germany was the linear accelerator that
Wideroe built for his PhD thesis (Chapter~4)\footnote{All references to chapters,
  sections, and figures refer to ``Beams'' mentioned in footnote 1.}
in 1927. Around 1936, rather than a linear accelerator, Gentner and
Bothe\footnote{Walther Bothe (1891-1957) received the Nobel Prize in Physics in 1954
  ``for the coincidence method and his discoveries made therewith.'' See also
  \url{https://www.nobelprize.org/prizes/physics/1954/}.} constructed
a van de Graaff accelerator (Figure~4.1) at the Kaiser-Wilhelm Institute of Medical
Science in Heidelberg. Once completed they used it for experiments in nuclear physics.
Assembling the machine had progressed very quickly and encouraged them to think about
a bigger accelerator---a cyclotron (Figure~4.4). Already in 1938 Wolfgang Gentner traveled
to Berkeley and established contact with the group of Lawrence, the inventor of the
cyclotron. As a result he obtained
blueprints for many parts needed to construct his cyclotron. By 1941---the Second
World War had already broken out---Gentner and Bothe had secured funding. Immediately
they started designing the new cyclotron and ordered the main components. During the
spring of 1943 the 70-ton magnet arrived and by the end of the year the cyclotron
produced deuterons with an energy of 12\,MeV. It operated into the 1970s.
\par
The cyclotron in Heidelberg was not the only one. Several others were planned in
Berlin, Leipzig, and Bonn.
\subsection*{Bonn}
In Bonn, a smaller 1.5\,MeV cyclotron was constructed at about the same time as Gentner's
machine in Heidelberg. Sadly it was destroyed during an allied bomb raid in October 1944.
\par
It took until 1952, after having learned about alternating-gradient focusing (Chapter~6),
that Wolfgang Paul\footnote{Wolfgang Paul (1913-1993) received the  Nobel Prize in Physics
  in 1989 ``for the development of the ion trap technique.'' See also
  \url{https://www.nobelprize.org/prizes/physics/1989/}.} applied for funds to build a
100\,MeV electron synchrotron (Chapter~8) in Bonn. Remarkably, the funding agency suggested
to increase the energy to 500\,MeV. After suitably adapting the application, the project
was approved  and resulted in the first alternating-gradient synchrotron in
Europe\footnote{For the history of this machine see: B. Mecking, {\em Das Bonner
    500\,MeV Elektronensynchrotron,} Bonn-IR-85-02 (in German).}. After designing the
machine with its 16\,m circumference throughout
the following years, often with the help of young students, construction commenced in 1955.
It took three years until all components were delivered and installed. But then a crisis
struck because it was realized that the magnetic field was unsuitable for stable operation.
The problem was eventually solved by sandwiching plywood sheets between magnets which
sufficiently changed the fields to allow stable operation. By the spring of 1958, first
beams were available and even synchrotron radiation was observed. In the following years
the beams were directed onto tungsten targets in order to produce ``bremsstrahlung'' which
are just highly energetic gamma rays. By directing these gamma rays onto protons or
deuterons and observing the scattering products (Chapter~2) much was learned about
the structure of atomic nuclei. The experimental program continued until 1984, when
the 500\,MeV synchrotron was dismantled to make space for ELSA, discussed below.
\par
Already a few years after the 500\,MeV synchrotron  produced beams, the need for a larger
electron synchrotron, reaching 2500\,MeV, became apparent. Towards the end of 1963, the
state government provided funds, such that design of the accelerator could
start\footnote{K. Althoff et al., {\em The 2.5\,GeV Electron Synchrotron of the University
    of Bonn,} Nuclear Instruments and Methods 61 (1968) 1.}. With a circumference of
almost 70\,m it became substantially larger than its smaller brethren. Four years later
the first electrons were injected from a 25\,MeV linear accelerator into the new synchrotron
and accelerated up to the design energy. After a while the beams could even be extracted
and directed towards a number of experimental stations, where they impinged on a variety
of targets, often containing protons or deuterons. The machine operated in this mode
until 1984, at which point the ELectron-Stretcher-Accelerator ELSA was constructed adjacent
to the larger synchrotron. 
\par
Synchrotrons provide a short burst of beams, typically 50 times per second, but many
experiments desire a continuous stream of electrons. One solution is to inject the short
burst into a larger so-called stretcher ring. In this ring one or several pulses
from the synchrotron are collected before the stored electrons are slowly extracted and
guided to experiments. The project already went under way in the early 1980s, but it took
until 1989 to complete the construction of ELSA\footnote{For an account of the history
  leading to construction of ELSA, see: W. Hillert, {\em The Bonn Electron Stretcher
    Accelerator ELSA: Past and future,} Eur. Phys. J. A 28, s01 (2006) 139.}
with its circumference of 164\,m and to connect it to the larger synchrotron that
was used as an injector. Once the electrons were inside ELSA, they could be slowly extracted,
just as was initially planned. Even accelerating the beams and extracting them at higher energies
became possible. Finally, beams could be accumulated in ELSA and stored for several hours
to provide synchrotron radiation. In this way, ELSA doubled up as a synchrotron light source
(Section~13.1).
\par
Much expertise was generated by constructing the smaller synchrotron, being the first of
its kind. And that competence came handy when a larger synchrotron was planned in Hamburg.
\subsection*{DESY}
By 1955 Germany had regained part of its sovereignty and was permitted to create
a ministry for nuclear energy. Moreover, this new ministry was willing to fund an
institute for research in particle physics; the creation of CERN in Geneva
the previous year served as a inspiration. Incidentally, at that time Hamburg
University offered a professorship to the US-based Austrian scientist Willibald
Jentschke, who accepted after being granted substantial funds to create a new
department, which served as the nucleus for what would later become
DESY\footnote{For a comprehensive account of DESY's first fifty years, see:
  E. Lohrmann, P. S\"oding, {\em Von schellen Teilchen und hellem Licht,}
  Wiley-VCH Weinheim, 2009 (in German). The second edition from 2013 is available
  online from \url{https://bib-pubdb1.desy.de/record/374686}.}.
\par
The type of accelerator was settled at a conference in Geneva the following year.
Instead of a proton synchrotron, which would be in direct competition with
the accelerator at CERN (Chapter~6), Jentschke decided to build an electron synchrotron
(Chapter~8) with the unprecedented energy of 6\,GeV, more than ten times the energy
of the first synchrotron in Bonn. During the following years negotiations between
federal and state agencies led to the constitution of the ``Deutsches Elektronen
SYnchrotron'' DESY in December 1959. An abandoned airfield in the suburb of
Bahrenfeld was selected as the site to house a 40\,MeV linear accelerator, used
as injector, and the 300\,m long synchrotron. It took until 1964 for the first
beams to be accelerated. Fifty times per second a new batch reached top energy,
was ejected from the synchrotron, and directed towards experimental stations.
\par
The first experiments pointed the electrons onto hydrogen-rich targets and analyzed
the scattering products in spectrometers, conceptually similar to Hofstadter's
experiments discussed in Chapter~7. A second group of experiments directed the
electrons onto a target where they produce gamma-ray photons that are then
impinged onto protons. Subsequently the photon-induced reactions and scattering
products are analyzed. A third group of experiments directed the electrons into
a bubble chamber (Figure~5.5), photographing the tracks, and analyzing the reactions
that show up. Right after beams became available, the copious synchrotron radiation
emitted by the 6\,GeV electrons was characterized and used for first experiments.
\par
\begin{figure}[tb]
\begin{center}
\includegraphics[width=0.99\textwidth]{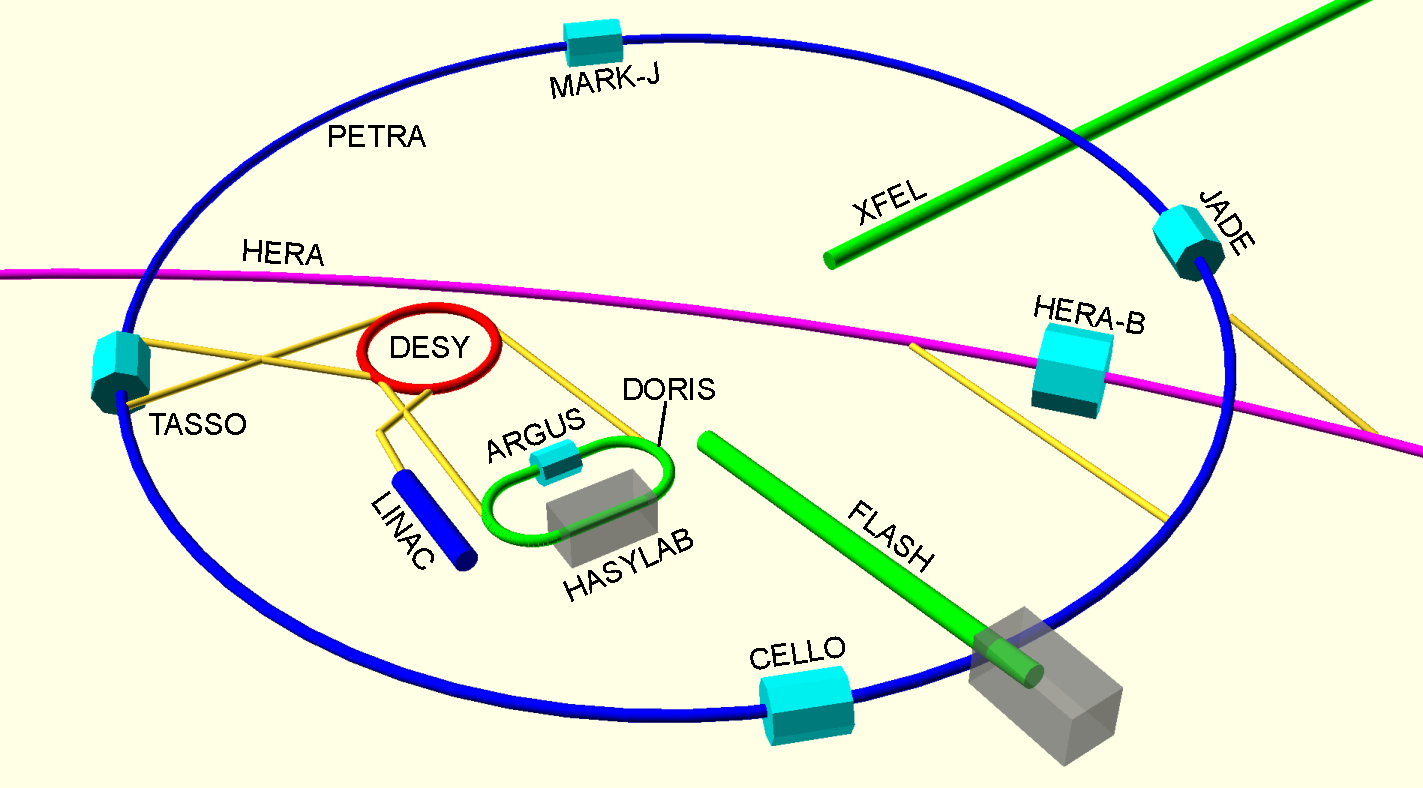}
\end{center}
\captionsetup{labelformat=empty}
\caption{\label{fig:desy}Sketch of the accelerators at DESY.}
\end{figure}
Already a few years after the synchrotron started to operate, ideas were discussed
to build a second accelerator in which electrons and positrons would circulate
in opposite directions and collide head-on (Figure 8.1). In this way much more energy
would become available for new particles to emerge from the fireball at the
collision points. There detectors
stood by to record the escaping particles. At the time it was deemed advantageous to
store electrons and positrons in separate rings, mounted on top of each other.
This new storage ring was given the name DORIS (derived from the German ``DOppel-RIng
Speicher''). It had a circumference of 288\,m and initially stored electrons and
positrons with energies of up to 3\,GeV.
\par
Towards the end of 1974, the two detectors PLUTO and DASP were placed in their
respective collision points. At just this time SLAC and Brookhaven discovered
the very sharp resonance made of charm and anticharm quark (Chapter~8), an
event later referred to as the ``November revolution.'' DORIS verified these
discoveries within a very short time and then successfully searched for mesons
consisting of one charm and one other quark. Later it engaged in the detailed analysis
of many bound states, an activity that was dubbed ``charmonium spectroscopy.''
Just as upgrading DORIS to higher energies was under way, the Upsilon as
a bound state of two bottom quarks was discovered at Fermilab in 1977 (Chapter~9).
Luckily the magnets were already somewhat over-designed from the beginning and
could handle the higher energies. Nevertheless, additional components and smart
ideas were asked for to eventually reach close to 5\,GeV per beam which made the
production of Upsilons possible. Observing their decay into muon and electron
pairs were important ingredients to determine the charge of the bottom quark.
DORIS then embarked on a campaign to determine other properties of the quarks
and the strong interaction.
\par
In order to consolidate operating at higher energies DORIS was upgraded in 1982
to reemerge as DORIS II. It now stored electrons and positrons in a single
ring, rather than the ``Doppel-ring'' that gave DORIS its name. The higher
energies of 5.6\,GeV per beam allowed the two new detectors, ARGUS and Crystal
Ball, to investigate pairs of $B$-mesons and their antiparticles $\bar B$, each
containing a bottom quark and one other quark. This led to the first observation
of $B$-$\bar B$-mixing, a process closely linked to the prevalence of matter
over antimatter in our universe that is referred to as CP-violation (Chapter~10). 
\par
All the time, the number of synchrotron-radiation users increased. A large
laboratory building housing their experiments was completed by 1982 and was
dubbed ``Hasylab,'' derived from ``HAmburger SYnchrotron LABoratory.'' Increasing
demands led to yet another upgrade to DORIS III in 1991, now equipped with an
outward-bulging no-longer-straight section. There five wigglers and undulators
(Section 13.1) were installed to increase the flux and the quality of the emitted synchrotron
radiation. By that time most particle physics activities concentrated on the
next-generation colliders PETRA and HERA such that from 1993 until 2012 DORIS
operated solely to produce synchrotron radiation. Research highlights include
Ada Yonath's studies\footnote{Ada Yonath (b. 1939) received the Nobel prize
  in Chemistry ``for studies of the structure and function of the ribosome.''
See also \url{https://www.nobelprize.org/prizes/chemistry/2009/}.}
of ribosomes, the engines that produce proteins in cells.
\par
Already by the time DORIS was starting up, discussions about the next generation
of accelerators were picking up speed and converged towards the electron-positron
collider PETRA (from the German ``Positron-Electron Tandem Ring\-beschleuniger\-Anlage'').
With a circumference of 2.3\,km it was the largest ring that would fit onto the DESY
site. It should reach more than 20\,GeV per beam and provide four collision points
for experiments housing the detectors TASSO, MARK J, JADE, and PLUTO; the latter was
later replaced by CELLO. Approval came in 1975 and less than three years later the
beams were colliding. The hope to find the sixth quark, called top, remained
unfulfilled. It turned out to be much too heavy for PETRA and other rings of similar
size. Instead, the highlight of research on PETRA was the discovery of gluons
through the observation of three-jet events (Figure~8.7), already mentioned towards
the end of Chapter~8. Many careful measurements revealed intricate details
of subatomic interactions and contributed to establishing quantum
chromodynamics as a valid theory of the strong interaction (Chapter~7). Moreover,
many aspects of the electroweak interaction (Chapter~6) were explored. The experimental
program continued until the end of 1986. At that time PETRA was needed as injector
for an even larger ring, the ``Hadron-Electron Ring Accelerator'' HERA. 
\par
After plans to add a proton ring inside the PETRA tunnel were abandoned,
a proposal to build a new 6.3\,km long underground tunnel was put forward. This
tunnel would house HERA comprising a 820\,GeV proton and a 30\,GeV electron ring.
Head-on collisions of electrons and protons at these energies would allow
the experimenters to probe the inside of the proton much more accurately than was
possible in the deep-inelastic scattering experiments at SLAC (Chapter~8). After
a period of negotiations the project was approved in 1984 and construction
could start. A large tunnel-boring machine, similar to the one that dug the tunnel
of LEP (Chapter~10) drilled the 6.3\,km long tunnel 10\,m to 20\,m below the surface.
The DESY synchrotron, rebuilt and called DESY II, as well as PETRA were used
to accelerate the
electrons. Accelerating protons, on the other hand, was new for DESY and a second
synchrotron for protons, called DESY III, was installed alongside the venerable
electron synchrotron. Additional components
were added to PETRA, now called PETRA II, that allowed it to accelerate protons.
The electron ring of HERA is fairly conventional, but the magnets for the
proton ring had to be superconducting in order to fit the high-energy beams into
the tunnel, just as the Tevatron needed superconducting magnets at Fermilab (Chapter~9).
After meticulous
development work at DESY the manufacture of the 750 up to 9\,m long magnets was
transferred to industry, half in Germany, the other half in Italy. The magnets
were cooled to low temperatures by the, at the time,
largest Helium liquefaction plant in Europe. After installing and testing many
more components, HERA was inaugurated towards the end of 1990. Commissioning all
subsystems, including the detectors H1 and ZEUS, took more time but by 1992
first collisions were observed. Later even two additional detectors, HERA-B and
HERMES, were installed.
\par
The detectors extended the deep-inelastic scattering results, pioneered at the
SLAC linac, to much smaller distances inside the proton, thereby validating the
standard model. In particular, the momentum distribution of the quarks inside
protons was measured with unprecedented accuracy. Moreover, there appeared to
be additional quark-antiquark pairs bouncing around inside protons. Even previously
unseen reactions that have their origin in the weak interaction were found; only a
single shower of hadrons appeared besides an unobservable neutrino. Its existence
was inferred by balancing the energy and the momentum of the observed ejectiles.
Experiments in HERA continued until
2007, when the accelerator was turned off, because DESY started to build the
European XFEL, a large free-electron laser (Section~13.2) to produce X-rays. With
HERA turned off, PETRA was no longer needed as injector and one octant was be
refurbished to house undulators that generate copious amounts of high-quality
synchrotron radiation. The rebuilt machine was called PETRA III. As of 2024,
even a further upgrade, called PETRA IV, to improve the amount and quality of
the emitted radiation, is actively discussed.
\par
The work towards the XFEL started from designing a linear electron-positron collider
(Chapter~12) called TESLA, that would use superconducting accelerating structures
(Chapter~10). In order to broaden the number of users beyond the particle physics
community, an X-ray free-electron laser was added to the proposal; it would serve
the community of synchrotron radiation users. In order to test the concept---both
the accelerating structures and the free-electron laser---a 100\,m long
superconducting linac accelerated beams to about 250\,MeV and passed them through
a 13\,m long undulator magnet. With all parameters tuned just right, the
machine would emit a high-intensity free-electron laser pulse (Section~13.2).
This Tesla-Test Facility (TTF) was constructed on the DESY site and produced very
short-wavelength light pulses in 2000. Shortly afterwards the machine, renamed
FLASH\footnote{For a comprehensive review see: J. Rossbach, J. Schneider,
  and W. Wurth, {\em 10 years of pioneering X-ray science at the Free-Electron Laser
    FLASH at DESY,} Physics Reports 808 (2019) 1.} (for ``Free-Electron LASer in
Hamburg''), was rebuilt to increase the energy to 1000\,MeV. Moreover, a new undulator,
now 30\,m long, led to first lasing at much shorter wavelengths by 2004. Continued
upgrades and an additional undulator have vastly enhanced the capabilities of FLASH.
In one of the many experimental highlights the scattered light from a microscopic
sample was processed to reconstruct the sample with high fidelity; a precursor to
determine the structure of many otherwise inaccessible biomolecules.
\par
But originally TTF was constructed to develop the superconducting accelerating
structures that were successfully used in TTF and FLASH. Despite this success
the application to build TESLA was turned down. Instead, based on the success of
the free-electron laser program, the funding agencies supported the construction
of a much larger version of FLASH, the European XFEL. About two thirds of this
3.2\,km long machine is filled with superconducting structures to reach beam
energies close to 20\,GeV. The last
third is filled with several-hundred-meter-long undulator magnets that produce
ultra-short and super-intense pulses of X-rays. Since the completion of the
facility in 2017, this radiation permits experimenters to investigate the
structure of proteins, make time-resolved measurements (``movies'') of chemical
reactions, and explore matter under extreme conditions.
\par
At DESY, a particle physics lab was slowly converted to one that pursues
synchrotron radiation research. In Berlin, on the other hand, a facility
dedicated to the generation of this radiation right from the start was constructed.
\subsection*{BESSY in Berlin}
By 1975, both the accelerators in Bonn and Hamburg were producing synchrotron radiation,
quasi on the side, but the demand was growing. As a consequence a review was conducted
and found that two storage rings, dedicated to the production of synchrotron radiation,
were needed. A larger one for the production of X-rays that turned out to be DORIS and
a smaller machine to generate radiation with lower energies. The formal decision to build
the latter came in 1979 and construction of BESSY (derived from ``Berliner ElektronenSpeicherring
Gesellschaft f\"ur SYnchrotronstrahlung'') soon commenced\footnote{D. Einfeld, G. M\"uhlhaupt,
  {\em Choice of the Principal Parameters and Lattice of BESSY, an 800\,MeV Dedicated Light
    Source,} Nuclear Instruments and Methods 172 (1980) 55.}. Less than three years later,
beams, accelerated
in a small pre-accelerator, and an injector synchrotron to 800\,MeV, were injected into
the BESSY storage ring having a circumference of 62\,m. Throughout the 1980s and 1990s
the beam quality and intensity were improved and with it the quality of the synchrotron radiation
that escaped the beam pipe through 35 windows. Highlights of the instrumental developments
are new types of undulator magnets\footnote{Of particular interest are
  so-called Apple undulators that allow to adjust the polarization of the emitted
  radiation. Today they are used in practically every synchrotron light source.}
(Section 13.1) and methods to ensure the quality of optical components that guide
and focus the synchrotron light on its way to the experimental stations. 
\par
Highlights of the experimental program are Peter Gr\"unberg's experiments on
what was later called giant magnetoresistance. This effect led the way to vastly
increased storage capacities of hard drives in our computers and was honored with
a Nobel Prize in Physics\footnote{Peter Gr\"unberg (1939-2018) received the Nobel Prize in
  Physics in 2007 ``for the discovery of giant magnetoresistance.'' See also
  \url{https://www.nobelprize.org/prizes/physics/2007/}.}. Surface physics
and catalysis are other fields of endeavor that were pursued at BESSY. These
phenomena are tremendously important to facilitate difficult chemical reactions,
such as converting toxic gases in car exhausts to more benign chemicals. The
high impact of this research contributed to Gerhard Ertl's Nobel Prize in
Chemistry\footnote{Gerhard Ertl (b.1936) received the Nobel Prize in Chemistry in
  2007 ``for his studies of chemical processes on solid surfaces.'' See also
\url{https://www.nobelprize.org/prizes/chemistry/2007/}.}.  
\par
The distribution of wavelengths---the spectrum---and the intensity of the emitted
radiation can be calculated from first principles and only depends on very few beam
parameters. Since those quantities can be very accurately measured, the emitted
radiation is known with high confidence and can be used to calibrate other measurement
devices, such as spectrometers. These activities were driven under the auspices of
the German authority for measurements, abbreviated PTB\footnote{PTB is a acronym,
  derived from ``Physikalisch Technische Bundesanstalt.''}. In particular, they were responsible
for the calibration of instruments for two space-telescopes, SOHO and Chandra. The
accelerator operated until 1999 and was then packed up and shipped to Jordan, where
it was reassembled and now serves as injector for SESAME, the first synchrotron light
source in the middle-east.
\par
A few years after the first BESSY storage ring started to produce light in the early
1980s, proposals for several so-called third-generation light sources with vastly
improved quality and intensities were discussed world-wide. Such an accelerator would
be several times larger than BESSY and would not fit within the limits of the original
site in West-Berlin. With the fall of the Berlin wall in 1989, however, new options appeared
and a site in former East-Berlin was chosen to become the home of BESSY-II\footnote{For
  more background on BESSY-II, see the brochure: {\em 25 Years of BESSY-II: A Success Story in Berlin-Adlershof,}
  special edition of ``Lichtblick'' from September 2023; online available from the following web page
  \url{https://www.helmholtz.de/en/newsroom/article/25-years-of-bessy-ii-light-source-for-the-good-of-society/}.}.
Construction
started in 1993 and five years later an injector synchrotron filled the new BESSY-II
storage ring, having a circumference of 240\,m and operating at 1.7\,GeV for the first
time. A year later first photons, produced in special undulator magnets but also
in the magnets that keep the beam on its circular path, were served to users. The
photons follow a straight path and escape from the beam pipe to travel via more than
forty so-called synchrotron beam lines towards the experimental stations.
\par
With higher beam energies, radiation in the X-ray region could be produced. It was
used to analyze the structure of materials and of proteins. Even the structure of proteins
of the Covid virus were determined at BESSY-II. A further very active field of research
addresses materials for new batteries and for high-temperature superconductors. Both
aspects are particularly relevant to improve the efficiency of energy utilization.
Even the PTB continues to calibrate instruments with the help of BESSY-II,
especially at higher photon energies. The reduced availability of lower-energy photons 
caused the PTB to construct a second smaller ring, the ``Metrology Light Source'' (MLS)
adjacent to the BESSY-II site.
\par
Construction of the MLS\footnote{G. Brandt et al., {\em The Metrology Light Source -- The
    new dedicated electron storage ring of PTB}, Nuclear Instruments and Methods B258 (2007) 445.}
started in 2004 and first beams became available in 2008,
predominantly for metrology applications, such as calibrating spectrometers but
also other sources of light used in industrial processes or in scientific applications.
The MLS has a circumference of 48\,m and stores electron beams with energies between
105\,MeV and 630\,MeV. Remarkably, the MLS reproducibly stores beam intensities ranging
from a single electron to several ten billions of electrons, with the intensity of
the emitted radiation covering the corresponding
range. Moreover, the synchrotron radiation covers the far infrared to the extreme
ultraviolet. It thus extends from far below the visible spectrum to significantly
above. Since the properties of the emitted radiation can be calculated from first
principles, the MLS serves as a ``standard candle'' to calibrate other sources or
spectrometers.
\par
Bessy and DESY were not the only laboratories producing synchrotron radiation. Towards
the end of the millennium, mid-size electron accelerators appeared in several other
places, among them Dortmund.
\subsection*{Dortmund}
%
Following the decision on the European level to locate the European Synchrotron
Radiation Facility in Grenoble instead of Dortmund, negotiations on the regional
level succeeded to secure funds for building a smaller accelerator, the ``Dortmund
ELectron Test Accelerator'' DELTA\footnote{D. N\"olle et al., {\em DELTA a new
    storage-ring-fel facility at the University of Dortmund,} Nuclear Instruments
  and Methods A296 (1990) 263.}. Construction started in 1990 and and by the second
half of the 1990s first beams were accelerated in a linac and a synchrotron to 1.5\,GeV
before they were stored in the DELTA storage ring with a circumference of 115\,m. The
declared purpose of DELTA was developing and testing of components for larger
facilities, preparing experiments with synchrotron radiation, educating a new
generation of accelerator builders\footnote{I am one of the substantial number of
  graduates that emerged from DELTA and populate accelerator labs all over Europe.},
and a first free-electron laser experiment called FELICITA\footnote{D. N\"olle et al.,
  {\em First lasing of the FELICITA I FEL at DELTA}, Nuclear Instruments and Methods
  A445 (2000) 128. FELICITA is an acronym for ``Free-Electron Laser In a Circular Test
  Accelerator.'' It also happens to be the middle name of my wife; what a coincidence.}.
\par
After the free-electron laser experiment was successfully completed the demand for
synchrotron radiation from the user community increased. As a consequence the originally
foreseen program was ramped down while the production of photons to serve users
was ramped up. The increased demand was then satisfied by installing additional wigglers
and undulators to increase the experimental capabilities to provide an increasing
number of users with synchrotron radiation.
%
\subsection*{Karlsruhe}
Before building the first linear accelerator (Chapter~4), Rolf Wideroe had already
conceived the betatron (Chapter~4) while he was a student in Karlsruhe, though he
was advised not to pursue the idea further. Much later, in the 1980s, ideas of building
a much larger accelerator emerged and a storage ring for industrial applications of
synchrotron radiation was proposed. This proposal was refined throughout the 1990s and
led to the construction of ANKA (from ``\AA Ngstr\"omquelle KArlsruhe''), an
electron storage ring with a circumference of 110\,m and a beam energy of
2.5\,GeV\footnote{G. Buth et al., {\em Status of the 2.5\,GeV Light Source ANKA},
  Proceedings of EPAC 1998 in Stockholm, p. 635.}.
It stored the first beams in 1999. The high beam energy and the large stored
currents cause the emission of very intense X-rays that are directed to multiple experimental
stations. Moreover, industry could simply purchase beam time from ANKA. This differs
remarkably from the way how beam time is allocated in accelerators that operate for pure
science. There it is assigned in a beauty contest---the scientifically most attractive
experiments get beam time.
\par
One of the prime uses of the radiation from ANKA and of particular interest for the
semiconductor industry is the creation of microscopic mechanical structures---precursors
of the accelerometers and gyroscopes found in modern phones---in a process called
LIGA\footnote{LIGA is the acronym derived from the German phrase ``Lithographie, Galvanoforming,
  und Abformung.''}, which is a type of X-ray lithography. The irradiation transfers
the geometrical pattern of a microscopic mask onto a substrate that is covered by a
protective film. In a second step the irradiated sections of the film are removed, and
in a third step an acid etches the pattern into the substrate, for example, silicon.
In another type of experiment the X-rays determine the structure of samples, including
proteins, on atomic scales. The wavelength of the radiation that is transmitted
to the samples can be adjusted in so-called monochromators to excite
deep-lying energy levels of atoms. Since these levels are specific to the type of
atom, very low concentrations of impurities in materials can be identified. 
\par
Following a restructuring of the laboratory in 2015, access to external users was
limited and this was reflected by changing the name from ANKA to KARA (for ``KArlsruhe
Research Accelerator'') and mainly using it for in-house developments\footnote{A. Papash
  et al., {\em Different operation regimes at the KIT Storage Ring KARA}, Proceedings
  of IPAC 2021 in Campinas, p. 163.}. After the reorganization
KARA was increasingly used to develop advanced beam diagnostic equipment such as
ultra-fast cameras to explore the dynamics of charged particle beams on previously
inaccessible time scales. Apart from developing other super-fast electronics components,
new technologies, among them superconductivity, were used to construct undulators
(Section~13.1) and other very compact magnets, important for future accelerators.
\par
In order to  complement the very short-wavelength radiation from KARA, in 2017 the 12\,m
long linear accelerator FLUTE\footnote{M. Nasse et al., {\em FLUTE: A versatile linac-based
    THz source,} Rev. Sci. Instrum. 84 (2013) 022705.} (derived from the German ``Ferninfrarot Linac Und Test
Experiment'') started to operate. It produces radiation with much longer wavelengths
than KARA. FLUTE accelerates electron beams to 50\,MeV, compresses them to a very short
length, and then passes them through a thin foil, where ultra-short pulses
of so-called THz radiation are produced. This radiation is immensely useful to study the
dynamics of chemical bonds that bind atoms into molecules, a process of great importance
in biology, as well as material and life sciences. 
\subsection*{Dresden}
Already in 1995 ideas were discussed to build a linear accelerator dubbed
ELBE\footnote{ELBE is an acronym derived from ``Electron Linear accelerator with
  high Brilliance and low Emittance,'' certainly inspired by the river Elbe that
  flows through Dresden.} in Rossendorf near Dresden that would
deliver a continuous stream of electrons for experiments in nuclear and radiation
physics\footnote{A. B\"uchner et al., {\em The ELBE-Project at Dresden-Rossendorf},
  Proceeedings of EPAC 2000 in Vienna, p. 732.}. To keep operating costs for this
linac reasonable the accelerator had to be
superconducting. Moreover, high demands on the beam quality stimulated the development
of high-performing electron sources. By 2001 the first 6\,m long accelerator stage
delivered beams up to 20\,MeV, later a second stage was added and brought the
energy to 40\,MeV. A system of beam transport lines directs the electrons to
several experimental areas\footnote{M. Helm et al., {\em The ELBE infrared and THz
    facility at Helmholtz-Zentrum Dresden-Rossendorf}, Eur. Phys. J. Plus (2023) 138.}.
\par
There the electrons are used to drive two free-electron lasers producing far-infrared
radiation that is used to probe material-science samples. The radiation is often
used in so-called pump-probe experiments. First a sample is excited with a pulse from
the free-electron laser and then a delayed second pulse from a conventional laser
probes the sample. In this way, the temporal evolution of very fast processes can
be investigated.
\par
A specially developed electron source produces electron bunches that are shorter
than the wavelength of the radiation emitted in a short undulator magnet. This
causes a dramatically increased emission, because all electrons radiate in phase.
The enhanced radiation level makes it possible to explore the properties of graphene,
among other things. 
\par
The electrons are used for many other purposes. They are crashed into targets and
produce very high-energy bremsstrahlung to probe the energy levels of nuclei.
Moreover, they are used to produce neutrons by crashing the beam into a liquid-lead
target. Even positrons can be produced. They are used to probe material defects in
metals and in semiconductors.
\subsection*{Darmstadt}
Inspired by Hofstadter's electron-scattering experiments in Stanford (Chapter~7),
the nuclear physics department in Darmstadt purchased a 60\,MeV linac from Varian
Associates (Chapter~7). By 1962 the 10\,m long linac, called DALINAC, started to
produce beams that impinged on a variety of targets\footnote{F. Gudden et al.,
  {\em Eine Anordnung f\"ur Experimente zur Elektronenstreuung unterhalb 60\,MeV},
  Zitschrift f\"ur Physik 181 (1964) 453 (in German).}. The escaping particles
were analyzed in spectrometers not unlike the one shown on Figure~7.4. In particular,
details of the nuclear structure were determined by inelastically scattering electrons
(middle image in Figure~7.3) from a large variety of nuclei, all the way up to uranium.
The linac operated well into the 1980s at which time a more powerful linac with better
beam quality was discussed.
\par
The new accelerator was called S-DALINAC and consisted of a 16\,m long linac based
on superconducting radio-frequency structures (Chapter~10). Two recirculation
arcs that allowed reusing the same linac three times, at which point the electrons
reached 130\,MeV and were directed to experimental areas. Instead of a pulsed beam,
the new linac provided a continuous stream of electrons. First beams were accelerated
in 1987 and henceforth used in nuclear physics experiments. Early on it was realized
that the new linac could be used for a free-electron laser (Section 13.2) as
well\footnote{V. Aab et al., {\em The Darmstadt near-infrared free-electron laser
    project}, Nuclear Instruments and Methods A272 (1988) 53.}.
After installing additional hardware, the laser started to produce the first photons
in 1996. These photons are routed to a separate room, dedicated to optical experiments. 
\par
Apart from experiments for nuclear physics and with the photons from the free-electron
laser, S-DALINAC serves as a test facility to explore advanced manipulations with
the beams themselves. In one of these experiments, the energy of the beam is recovered
inside the structures that were used to accelerate it in the first place. The
original beam is discarded, but its energy is reused to accelerate new beams, thus
the machine becomes a so-called energy-recovery linac.
\par
Not only electrons can be used in nuclear physics experiments, also accelerated beams
of heavy ions provide suitable projectiles to explore the inner workings of nuclear
matter. Close to Darmstadt, in a nearby village, a new lab appeared in the late 1960s.
\subsection*{GSI and FAIR}
During the 1960s accelerators for heavy ions were discussed\footnote{For an early account see:
  K. Blasche, H Prange, {\em Die GSI in Darmstadt (I),} Physikalische Bl\"atter 33 (1977) 249
  (in German). Online available from \url{https://doi.org/10.1002/phbl.19770330602}. The
  GSI website at \url{https://www.gsi.de/en/about_us/} provides a wealth of more up-to-date
  information.} as tools to analyze nuclear
matter with methods that were complementary to those of high-energy proton or electron
accelerators that were used to look inside protons and other sub-nuclear particles at the
time. Instead, the new facility would focus on the collective behavior of large numbers of
protons and neutrons inside heavier nuclei. As a consequence of these discussions, the
``Gesellschaft f\"ur SchwerIonenforschung'' (GSI) was founded in 1969 and the construction
of the approximately 120\,m long linear accelerator, called UNILAC, soon started. It
was based on drift tubes developed by Wideroe (Figure~4.2) and Alvarez (Figure~5.3).
However, with a diameter of 1.2\,m the vacuum vessel is much larger. By 1975 it accelerated
beams up to energies of 11\,MeV per nucleon. The total energy of uranium with 238
nucleons (92 protons and 146 neutrons) is therefore in the GeV range. At these energies
the speed of the uranium ions is about 16\,\% of the speed of light. Soon experiments
in two experimental areas commenced by impinging the heavy-ion beams onto targets.
Detectors then analyze whatever scattering products escape from the point of impact.
\par
Of particular interest is the distribution of fragments with different numbers of
protons and neutrons (isotopes) that are created in the collision, because it shows
how the nucleons are organized inside the nucleus. Thereby it probes the organization
of nucleons into shells\footnote{Maria Goeppert-Mayer (1906-1975) and Hans Jensen
  (1907-1973) received the Nobel Prize for Physics in 1963 ``for their discoveries
  concerning nuclear shell structure.'' See also \url{https://www.nobelprize.org/prizes/physics/1963/}.}.
These experiments help us to better understand
the stability of nuclei, especially in extreme situations as they occur in very heavy nuclei.
Some of the lighter isotopes feature in the creation of heavier elements inside stars
and supernovae, a process called ``nucleosynthesis''.  In particular, measuring their
masses and lifetimes helps us understand the relative abundance of different elements
in our universe.
\par
The heaviest naturally occurring element on the periodic table is uranium with atomic
number 92. Since the 1940s heavier elements, so-called transuranium elements, were created in
nuclear reactors and in accelerators. Once the UNILAC was up and running, it joined
the search for previously unseen heavier elements. And indeed, between 1981 and 1984, the
three elements with atomic numbers 107, 108, and 109\footnote{The transuranium elements
  107 to 109 are nowadays called bohrium, hassium, and meitnerium, respectively.} were
discovered at GSI. Bohrium with atomic number 107, for example, was created by slamming
chromium ions into a bismuth target and then filtering the debris to select particles
with a specific mass using an elaborate setup of magnetic and electric fields, called SHIP,
short for ``Separator for Heavy Ion Products.'' Other combinations of beams and targets
were used to create the other new elements. And about ten years later, between 1994
and 1996, the team at GSI repeated their success and found elements 110, 111, and
112\footnote{The transuranium elements 110 to 112 are nowadays called darmstadtium,
roentgenium, and copernicium, respectively.}.
In the decade between the element discoveries, many other experiments were performed and,
most importantly, the facility was upgraded by adding a synchrotron and a storage ring.
\begin{figure}[tb]
\begin{center}
\includegraphics[width=0.99\textwidth]{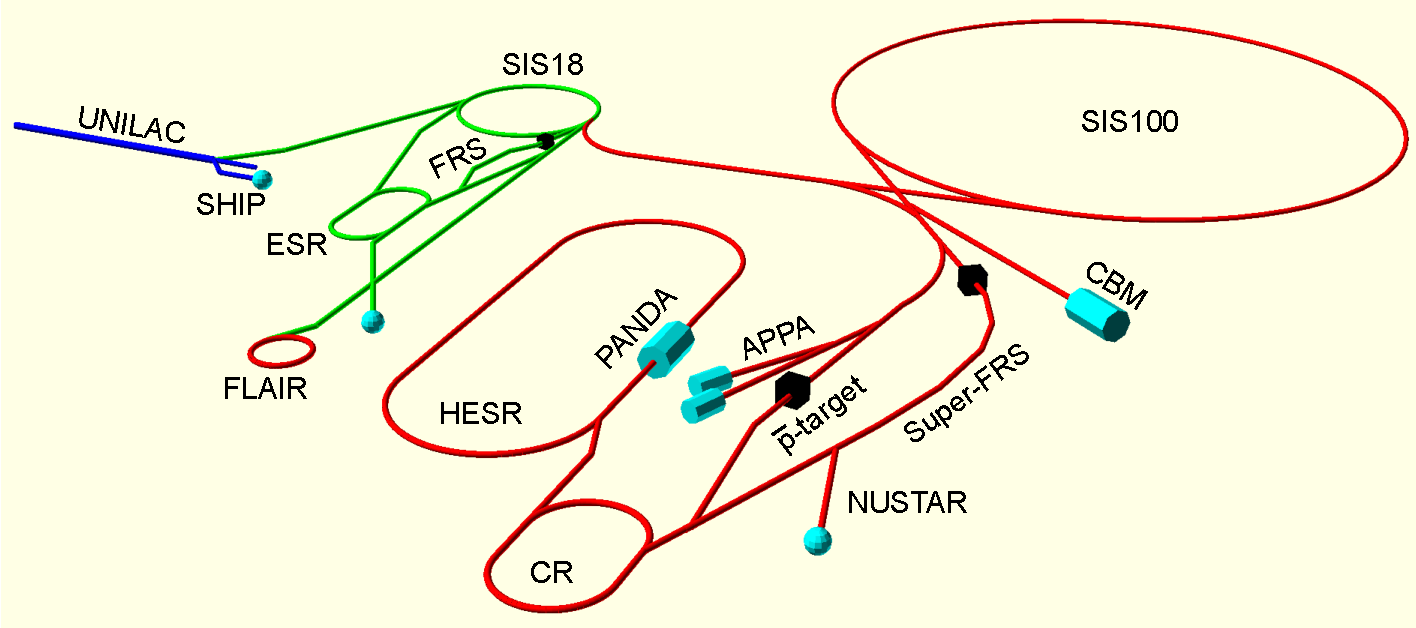}
\end{center}
\captionsetup{labelformat=empty}
\caption{\label{fig:fair}Sketch of GSI and the FAIR facility.}
\end{figure}
\par
Already in the early 1980s, different options to increase the energy of the beams were
discussed and finally converged on a synchrotron (Chapter~5), for technical reasons
called SIS18, and the ESR, short for ``Experimentier SpeicherRing.'' SIS18 has a
circumference of 216\,m and increases the energy of beams by large factors. At the
same time their speed increases from 16\,\% to 90\,\% of the speed of light. The
accelerated beams are then directed onto a target and the debris is filtered by
a special beam line consisting of magnets and thin absorbers, called FRS, short for
``FRagment Separator.'' Only one specific nucleus containing a particular number
of protons and neutrons (a specific isotope) emerges at the exit of FRS and is
injected into the ESR with half the circumference of SIS18. In the ESR a stochastic
cooling system (Figure~9.4) improves the beam quality of the injected ions.
Additionally, an electron cooler superimposes a high-quality electron beam with
the stored ion beam. This allows the ions to bounce off the electrons and thereby
lose some of their random motion. This process then slowly ``cools'' the ions and makes it possible
to measure their revolution frequency with very high precision from which the
mass of the injected isotope can be derived. This method was used to analyze a
wide variety of isotopes thus improving our understanding of nuclear matter.
\par
Since the early 1990s, GSI explored the use of ion beams to treat cancer.
Especially carbon ions showed a distinct advantage over conventional therapy
with X-rays or electrons, because the carbon ions cause much more damage in
deep-seated tumors while causing less harm in surrounding healthy tissue.
The depth of the tumor-damage can be adjusted by adjusting the energy of the
SIS18 synchrotron. The extension of the tumor in the sideways directions
is covered by raster-scanning the ion beam with the help of magnets. During
the process that destroys a tumor, positrons are created that annihilate
with nearby electrons and give rise to two characteristic X-ray photons that
reveal their point of origin. In this positron-emission tomographic (PET)
process the deposited dose is verified while the patient is treated. The pilot
project at GSI was such a success that a similar facility comprising synchrotron
and raster-scanning target system was built in Heidelberg and treats
patients from 2009 onwards. A few years later similar systems started
operating in Marburg, Shanghai (China), Pavia (Italy), and Vienna (Austria).
\par
Once SIS18 and the ESR were in operation in the mid-1990s, discussions started
about the future needs of the nuclear physics community. It came up with a proposal
for the ``Facility for Antiproton and Ion Research'' (FAIR)\footnote{See the online
  presentation at \url{https://fair-center.eu/} for more information.}. The funding
agencies recommended to jointly construct the facility with international partners.
By 2010, all partners signed an agreement and FAIR was officially founded. It still
took several years until construction began. Unfortunately it was delayed further
due to massive increases in the cost, the covid pandemic, and the Russian war in Ukraine.
By the time of writing construction, albeit on a reduced scope, is ongoing.
\par
For FAIR, all the existing accelerators at GSI are upgraded and will become part
of the injector complex for the large synchrotron SIS100. With a circumference of
1100\,m it is about five times larger than SIS18 and reaches about five times
higher energies. Beams extracted from SIS100 are guided to several experimental
areas; one of them is the Super-FRS, a large fragment separator to isolate exotic
isotopes. The ``Collector Ring'' CR collects these isotopes and makes detailed studies
of their properties possible, just as the ESR was used to analyze isotopes produced
in the FRS. In parallel to the Super-FRS, proton beams from SIS100 can be directed to
an antiproton-production ($\bar p$) target. Also the antiprotons are collected in the
CR. Once their beam quality is increased, they are directed to the ``High-Energy Storage
Ring'' (HESR) where they are stored for hours and collide with a hydrogen target inside
the PANDA detector.
\par
The physics program of FAIR comprises four topics, labeled NUSTAR, PAN\-DA, CBM, and
APPA\footnote{These names are acronyms. NUSTAR stands for ``Nuclear Structure,
  Astrophysics and Reactions,'' PANDA stands for `` antiProton ANnihilation at
  DArmstadt,'' CBM stands for ``Compressed Baryonic Matter,'' and APPA stands for
  ``Atomic, Plasma Physics and Applications.''}.
NUSTAR is focused on exploiting the Super-FRS and determines the masses and
lifetimes of rare isotopes that play an important role for the creation of heavier
elements in stars and during supernova explosions. PANDA explores the spectrum of
hadrons that include a charm quark and tries to find bound states of gluons,
so-called glueballs. Since the mass of the quarks alone does not account for the
mass of, for example, protons, PANDA tries to find explanations based on quantum
chromodynamics. CBM will crash the full-energy beam coming from SIS100 into a
target inside a dedicated cave. There conditions are recreated that resemble those
immediately after the Big Bang when the primordial quark-gluon plasma condensed
into the first hadrons. APPA explores how high-energy particles affect matter
relevant to life sciences and material sciences. For example the impinging particles
can form nanopores that serve as filters to separate smaller from larger compounds.
\par
Not far from the FAIR site, in Frankfurt and in Mainz, two institutes pursue
development work that directly contributes to GSI and FAIR. The ``Institute for Applied
Physics'' at Frankfurt University develops accelerating structures, especially for
the new linear accelerator HELIAC. It will take over the super-heavy element production
from the venerable UNILAC, once the latter is dedicated to FAIR. The ``Helmholtz
Institute'' in Mainz integrates the structures into HELIAC and works on a new
electron cooler for the HESR to produce high-quality beams for PANDA.
\subsection*{Mainz}
Long before the work towards FAIR, already since 1963, the university in Mainz operated
a 300\,MeV electron linear accelerator for experiments in nuclear physics. This
machine, however, did not work quite as well as was hoped for and, tongue in cheek, was
named MUELL (for ``Mainzer Universit\"ats ELectron Linac''). About a decade later, an
ingenious add-on, dubbed MALAISE (for ``MAinz Linear Accelerator Improving
SystEm'')\footnote{The German word ``Muell'', if spelled ``M\"ull,''
  translates to the English word ``garbage.'' Malaise, French for ``misery,'' is actually
  a misnomer, because performance of the linac improved substantially.}, improved the
beam quality significantly, which is essential when probing nuclear matter with electrons.
Experimental highlights comprise a precise measurement of the mixing of $Z^0$-bosons and
photons at low energies. This is a crucial parameter for the standard model, commonly
referred to as Weinberg angle.
\par
\begin{figure}[tb]
\begin{center}
\includegraphics[width=0.99\textwidth]{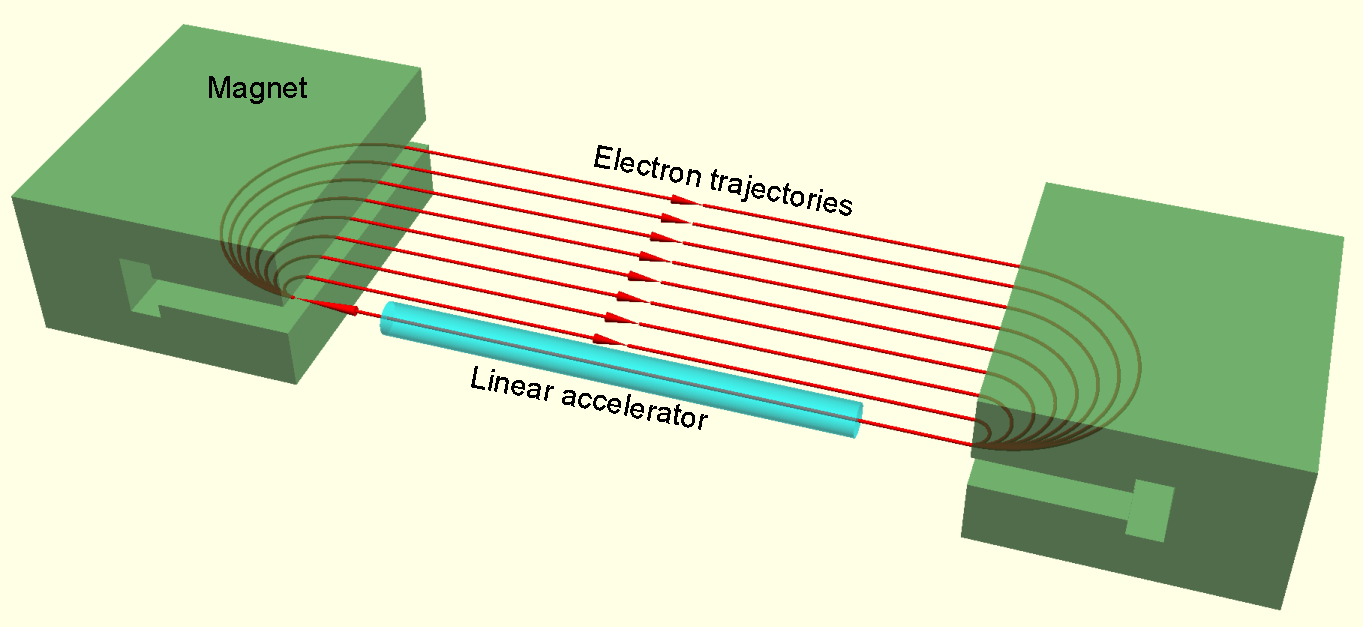}
\end{center}
\captionsetup{labelformat=empty}
\caption{\label{fig:micro}Sketch of a microtron.}
\end{figure}
But in the mid 1970s the experimenters demanded higher energies and better beam
quality, which would allow them to increase the precision of their experiments.
But most of all, they asked for a continuous stream of electrons. The accelerator
to provide these beams was dubbed MAMI\footnote{H. Herminghaus et al., {\em The
    Design of a Cascaded 800\,MeV Normal Conducting C.W. Race Track Microtron},
  Nuclear Instruments and Methods 138 (1976) 1.}, derived from ``MAinz MIcrotron.'' In
microtrons a short linear accelerator is reused multiple times by recirculating the
electron beam with magnets. This provides a much more compact way of reaching higher
energies than would be possible by traversing a longer linac only once. Nevertheless,
accelerating electrons from very low energies all the way to the full energy in
only one microtron would be technically impossible; therefore MAMI had to be constructed
as a cascade of three microtrons. The first one brought the energy up to 14\,MeV
and by 1983 the second microtron produced electrons with 213\,MeV. Constructing
the much larger third stage took more time, but by 1990 a continuous stream of
electrons with energies up to 855\,MeV was available. Most experiments
used these electrons to probe the structure of nuclei. And the continuously available
beam made it possible to probe very rare events. One just had to accumulate data
for a long time. Other experiments first crashed the electrons into a heavy target
to produce so-called bremsstrahlung composed of high-energy photons and then using
the photons to probe various nuclei.
\par
After running the facility for about a decade the experimentalists once again
requested higher beam energies. That stimulated the construction of a fourth
accelerator stage. Since the magnets for this large microtron would be too massive,
the accelerator builders came up with an ingenious proposal\footnote{A. Jankowiak,
  {\em The Mainz Microtron MAMI---Past and Future}, Eur. Phys. J A 28, s01 (2006) 149.}.
They proposed to split the large
magnets into smaller ones and adjust ancillary components such as the acceleration
system appropriately. By 2006 the fourth stage was up and running and produces
1500\,MeV beams ever since.
\par
Already the four MAMI stages could produce electrons with a well-defined orientation
of their spin, but certain measurements require much higher beam intensities to
achieve results in a reasonable time. To accommodate this type of experiment a new
accelerator, called MESA\footnote{MESA is an acronym derived from ``Mainz Energy recovering
  Superconducting Accelerator.'' See also \url{https://www.mesa.uni-mainz.de/eng/}.},
was proposed. In a second mode of operation, the
accelerated electrons would be passed through the same structures that accelerated
them, thereby recovering the energy of the beam and reusing it to accelerate a fresh
beam of electrons with better quality. This mode of operation makes much higher
beam intensities possible and thereby enables the experimenters to probe much
rarer processes than previously possible. In particular, searches for the hypothetical
dark matter (Chapter~14) are planned once MESA becomes operational in 2024.
\subsection*{Heidelberg}
Even before the previously mentioned cancer therapy center was built in Heidelberg,
an ion storage ring was constructed at the Max-Planck Institute for Nuclear Physics
in close collaboration with GSI. By 1988 the ``Test Storage Ring'' TSR\footnote{
  P. Baumann et al., {\em The Heidelberg Heavy Ion Test Storage Ring TSR},
  Nuclear Instruments and Methods A268 (1988) 531.} with a
circumference of 55\,m stored beams of ionized atoms and molecules. An
electron cooler, similar to the one installed in the ESR at GSI, was used to reduce
the beam sizes and allowed the experimenters to analyze the behavior of beams with
very high charge densities.
Moreover, the cooler was used as an electron target to investigate atomic and
molecular recombination processes. In these processes an atom picks up an
additional electron from the cooler. They are important to understand the creation
of complex molecules in outer space. Moreover, the accelerated and cooled ion beams
were used for high-resolution experiments in nuclear physics. The ring was operated
until 2012 when construction of a new storage ring in Heidelberg started.
\par
Rather than using magnetic fields to guide the beams, the new ring with a
circumference of 35\,m uses electric fields, which is advantageous for slow and
very heavy particles or molecules. Moreover, in order to permit very delicate
measurements, the ring is cooled down to extremely low temperatures and is therefore
dubbed CSR\footnote{R. von Hahn et al., {\em The cryogenic storage ring CSR},
  Rev. Sci. Instrum. 87, (2016) 063115.}, an acronym derived from ``Cryogenic
Storage Ring.'' Like in the earlier ring, an electron cooler serves as an electron
target for high-resolution recombination reactions.
\subsection*{J\"ulich}
Already 1968 the JULIC\footnote{L. Aldea et al., {\em Status of JULIC}, Proceedings
  of Cyclotrons 1981 in Caen, p. 103.} cyclotron in J\"ulich accelerated protons and other light ions
with energies up to 45\,MeV that were mostly used for nuclear physics experiments.
A number of transfer lines brought these beams to several different target stations,
among them a large detector called BIG KARL. Other stations were used to produce
radioactive isotopes for medical applications.
\par
Already in the early 1980s, ideas appeared to add a synchrotron to the facility that
would increase the energy of the protons substantially. In order to produce high-quality
beams, an electron cooler was planned, which gave the synchrotron the name
COSY\footnote{U. Bechstedt et al., {\em The cooler synchrotron COSY in J\"ulich},
  Nuclear Instruments and Methods B113 (1996) 26.}, for
``COoler SYnchrotron.'' It has a circumference of 184\,m and accelerates protons,
which could be prepared with their spins aligned, to energies of up to 2300\,MeV. 
Besides the electron cooler, elaborate stochastic cooling systems (Figure~9.4) improve
the quality of beams that are delivered to the experiments. Three of them are part
of the ring where the beam continuously interacts with a target such as a gas jet or a very
thin carbon filament. Later a stream of frozen hydrogen spheres, so-called pellets,
served as a target to explore nuclear interactions. Apart from using these internal
targets, the beam was extracted from the ring and directed to external detectors,
among them the venerable BIG KARL.
\par
The nuclear physics program continued until 2014, at which time the focus switched
to precision experiments to understand various aspects of CP-violation (Chapter~10).
In the ``Time Reversal Invariance at COSY'' (TRIC) experiment, a specially prepared
proton beam with spins aligned collided with a likewise prepared deuteron target.
A second experimental program,
called EDM, aims at measuring the ``Electric Dipole Moment'' of protons and deuterons
by observing the precession of the spin of beam particles in the presence of strong
electric fields. Any non-zero value would give strong indications about ``new
physics'' beyond the standard model. At the same time, COSY was used to develop the
next-generation accelerator technology needed for the FAIR facility in Darmstadt,
in particular the electron coolers and stochastic cooling systems.
\subsection*{And there is more}
Besides the larger accelerators covered in this report, there are many smaller ones.
Many hospitals operate small cyclotrons to produce radioactive isotopes that are used
for diagnostic purposes, such as positron-emission tomography (Section~13.4). They
also operate small, often less than one meter long electron linear accelerators
to irradiate tumors.
\par
Several universities operate electrostatic accelerators of the type devised by
van der Graaff (Figure~4.1). They are used for dating artefacts using the carbon-14
method (Section~13.5) and to analyze specimen by impinging the beams onto the sample.
Subsequently observing the emitted X-rays, specific to the material composition
of the specimen, allows the experimenters to identify even minute quantities of
constituents.
\par
Accelerating very heavy ions to a few hundred of kV implants the ions inside the
irradiated substrate, which is used in industry and academia. These machines
are up-to-date versions of Cockroft and Walton's accelerator (Figure~4.1) used
to split Lithium atoms in 1930.
\subsection*{Acknowledgements}
I benefitted greatly from helpful discussions with Christian Jung, Roger Ruber,
Detlev Schirmer, Wolfgang Hillert, Daniel Elsner, Andreas Lehrach, Kurt
Aulenbacher, and J\"org Rossbach.
%
\end{document}